\documentclass[rsc,twocolumn]{revtex4}

\usepackage{amsmath}
\usepackage{amsfonts}
\usepackage{amssymb}
\usepackage{graphicx}

\begin{document}

\title{Crack patterns over uneven substrates}
\author{Pawan Nandakishore}
\author{Lucas Goehring}
\email[]{lucas.goehring@ds.mpg.de}
\affiliation{Max Planck Institute for Dynamics and Self-Organization (MPIDS), 37077 G\"ottingen, Germany}

\date{\today}

\begin{abstract}
Cracks in thin layers are influenced by what lies beneath them. From buried craters to crocodile skin, crack patterns are found over an enormous range of length scales. Regardless of absolute size, their substrates can dramatically influence how cracks form, guiding them in some cases, or shielding regions from them in others. Here we investigate how a substrate's \textit{shape} affects the appearance of cracks above it, by preparing mud cracks over sinusoidally varying surfaces. We find that as the thickness of the cracking layer increases, the observed crack patterns change from wavy to ladder-like to isotropic.  Two order parameters are introduced to measure the relative alignment of these crack networks, and, along with Fourier methods, are used to characterise the transitions between crack pattern types.  Finally, we explain these results with a model, based on the Griffith criteria of fracture, that identifies the conditions for which straight or wavy cracks will be seen, and predicts how well-ordered the cracks will be.  Our metrics and results can be applied to any situation where connected networks of cracks are expected, or found.
\end{abstract} 

\maketitle

\section{Introduction}

From networks of cracks across the surfaces of planets \cite{Mcgill1992,Cooke2011,Freed2012,Blair2013} to artistic craquelure in pottery glazes and paintings \cite{Karpowicz1990,Bucklow1998,Pauchard2007}, or bespoke fracture textures in thin films \cite{Nam2012,Kim2013}, crack patterns naturally occur on a wide variety of scales.  They are even implicated in the formation of scales on crocodiles \cite{Milinkovitch2013}. All these structures represent the outcome of failure processes that occur in a brittle medium supported by an uneven substrate. Here we show how the shape of the substrate plays an important role in determining the geometry of the resulting cracks.  We will demonstrate how to use the substrate to template fractures, explore the conditions necessary to generate a variety of patterns, and introduce general metrics to quantify the orientational ordering of any set of connected cracks.

\begin{figure}
\centering
\includegraphics[width = 8.3 cm]{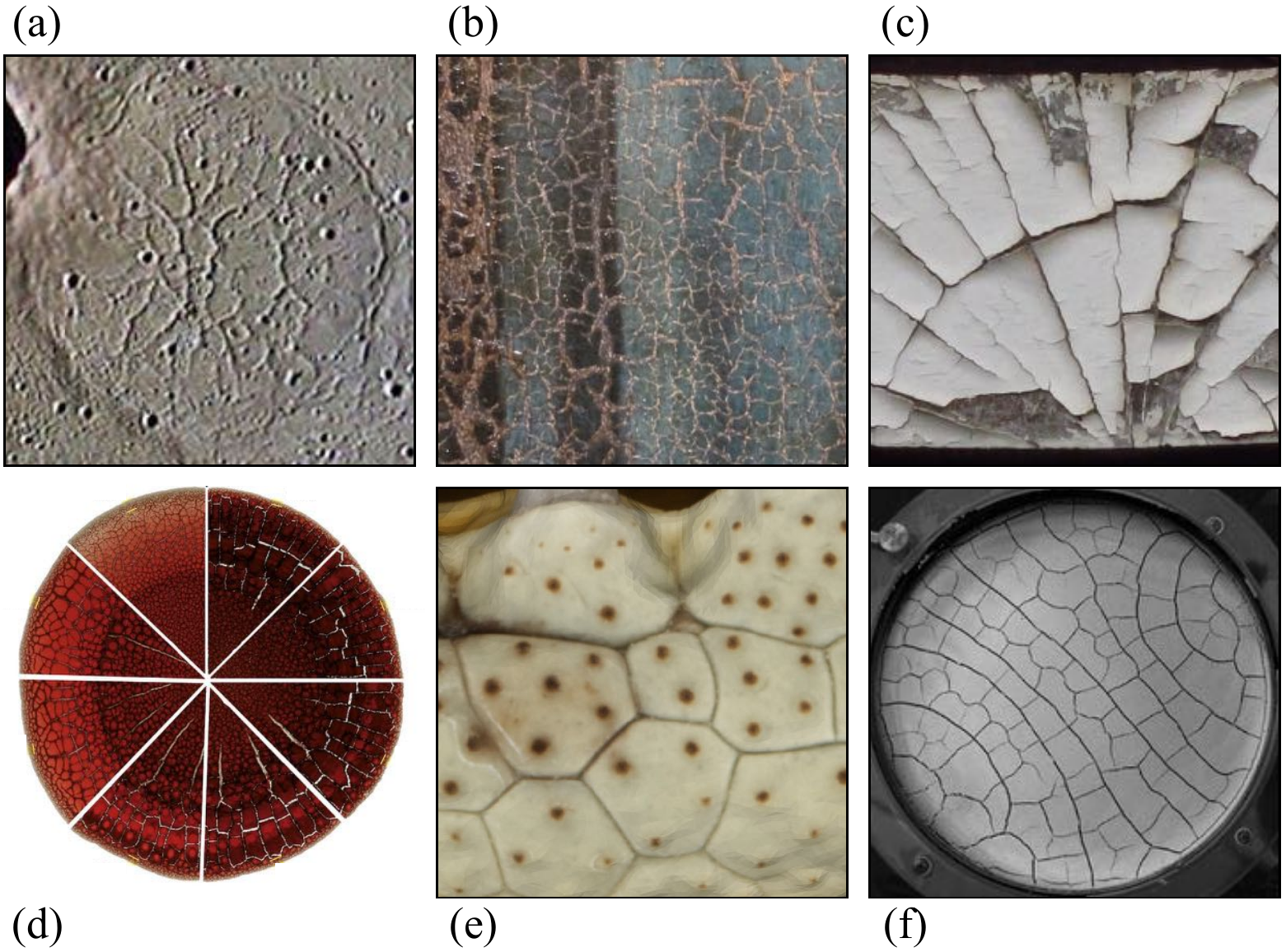}
\caption{Examples of crack networks, oriented by different effects. 
(a) Cracks in lava over a buried crater in Goethe Basin, Mercury \cite{NASA}. 
(b) Craquelure pattern of traditional oil-on-wood technique. 
(c) Paint cracks on a beam, reflecting the growth rings of the underlying wood. 
(d) Cracks in dried blood droplets, in a range of humidities, are affected by the way the droplet dries (courtesy of D. Brutin, adapted from \cite{Zeid2013}, Fig. 6). 
(e) Skin cracks on embryonic crocodiles are the result of differential growth of the skin layers, and form tiles that mature into scales (courtesy of M. Milinkovitch, adapted from \cite{Milinkovitch2013}, Fig. 1b).  
(f) Desiccation cracks in pastes are influenced by any vibration or flow prior to drying (courtesy of A. Nakahara, adapted from \cite{Nakayama2013}, Fig. 3b). 
} \label{IntroFig}
\end{figure}

To introduce the concerns of our work we will first briefly present three examples of cracks on the scales of kilometres, centimetres, and micrometers.  These motivational examples, and those shown additionally in Fig. \ref{IntroFig}, also aim to highlight the need for quantitative approaches to the analysis of fracture patterns.    
    
On large length scales, radial graben are associated with buried craters on Mercury \cite{Freed2012,Blair2013} and Mars \cite{Mcgill1992,Cooke2011}. These features are due to the repeated filling and cooling of lava in impact craters. The uneven stresses caused by the edge of the crater are believed to lead to wrinkle ridges above the crater rim   \cite{Blair2013}.  Contraction cracks also form within the interior of the crater, leading to a mixture of polygonal and circular graben patterns \cite{Freed2012,Blair2013}. The radial symmetry of the substrate imposes a characteristic structure to the such cracks, as in Fig. \ref{IntroFig}(a). 

On a more intermediate length scale, craquelure (Fig.~\ref{IntroFig}(b,c)) is the result of differential strains between a coating and its substrate \cite{Berger1990,Karpowicz1990,Bucklow1998,Pauchard2007}.  This may be due to tension in the framing of a painted canvas \cite{Karpowicz1990}, for example, or changes in the relative humidity at which a painting has been stored \cite{Berger1990}.   Alternatively, it may result from an artist's choice of a pottery glaze with a higher coefficient of thermal contraction than the clay it covers.  Such craquelure patterns are extremely hard to forge, and can thus aid in the attribution of paintings~\cite{Bucklow1998}.  They have been shown to be influenced by local painting style, choice of substrate, including, for paint on wood, even the species of wood used~\cite{Bucklow1998,Crisologo2011}.  The analysis of similar patterns, in dried blood (Fig. \ref{IntroFig}(d)), is also being developed as a forensic tool, for example to help determine the relative humidity when the blood was spilt \cite{Zeid2013,Zeid2013b}.

On the micrometer scale, recent attention has focussed on controlling cracks. For example, Nam et al. \cite{Nam2012} showed how to selectively initiate, refract, and stop both straight and wavy cracks in silicon nitride thin films, vapour-deposited on silicon wafers, by adding notches, cutting stairs, or introducing interfacial layers between the wafer and film.  Kim et al. \cite{Kim2013} instead looked at soft substrates, with crack formation guided by strategically placed notches along grooves; their work aims to \textit{control} failure in situations, such as flexible electronics, where it must be expected.   They find the most regular results when they design for crack spacings that are near the natural crack spacing of the brittle film.  

A common feature of all these situations is that the cracking layer is thin, compared to its lateral extent.   Nonetheless, the layer thickness imposes the natural length-scale of such problems, and the spacing of the cracks in each case reflects this thickness \cite{Allain1995,Bai2000,Shorlin2000,Kim2013}.  Here we study the model problem of contraction cracks in `thin' elastic layers adhered to rigid substrates and show how the shape of the substrate also plays an important role in determining the geometry of the crack pattern.   

For this end we choose the system of desiccation cracks in drying clays as a simple, yet representative, experiment.  As clays dry they tend to shrink, uniformly.   So-called channel cracks open in the film, to relieve its growing strain energy (see \textit{e.g.} \cite{Hutchinson1992,Beuth1992,Xia2000}).  The appropriate physical model for stress in a wet clay, poroelasticity, has an exact mathematical analogy with the thermoelastic model of how most solids shrink as they cool \cite{Biot1956,Norris1992,Goehring2015}.  In biology, comparable conditions can also result from the differential growth of layered tissues \cite{Couder2002,Laguna2008,Boudaoud2010}.  This is known to lead to the formation of scales on the heads of Nile crocodiles\cite{Milinkovitch2013} (Fig. \ref{IntroFig}(e)), for example, while a similar mechanism has been proposed to account for leaf venation patterns \cite{Couder2002,Laguna2008}.   

\section{Materials and Methods} 
 
\begin{figure}
\centering
\includegraphics[width =  8.3 cm]{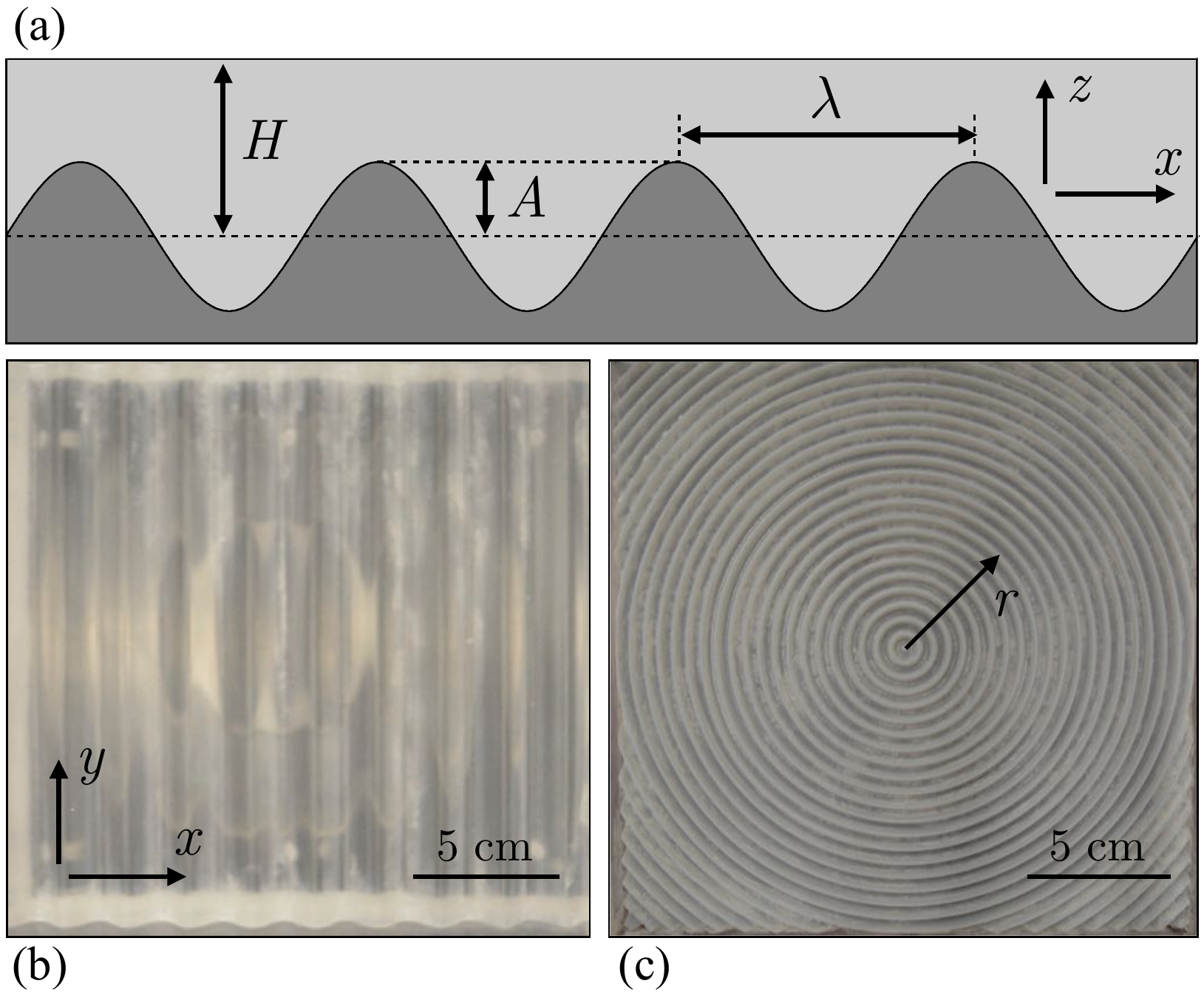}
\caption{Sketch of the drying experiment, where (a) a clay slurry of average initial thickness $H$ dries over a sinusoidally varying substrate, of amplitude $A$ and wavelength $\lambda$.  Both (b) linear and (c) radial wave plates were used as substrates.} \label{schematic}
\end{figure}

Briefly, bentonite clay was dried over rigid patterned substrates, where it cracked; a schematic summary of the experiment is given in Fig. \ref{schematic}(a).   Compared to the wet clay, the substrate is undeformable and unbreakable.

Each substrate consisted of a 20$\times$20 cm$^2$ plate.  Five sinusoidal plates were cut from acrylic blocks by computer-numerical-control (CNC) milling, with a resolution of 200-400 $\mu$m. These plates were corrugated; their surface $z$ varied sinusoidally along one direction, $x$, with a constant profile along the other, $y$, direction such that $z = A\sin(2\pi x/\lambda)$.  The wavelengths $\lambda$ and amplitudes $A$ of the plates are given in Table~\ref{allvals}.  Two pairs of plates had the same dimensionless ratios $a = A/\lambda$, in order to check the scale-independence of the fracture patterns.   Two additional plates were prepared by 3D printing from acrylic photopolymer, with a vertical resolution of 30 $\mu$m, and a contour resolution of 5 $\mu$m (4D Concepts).   These plates had radial wave patterns, with surface profiles (see Table~\ref{allvals}) that varied periodically in the radial direction, $r$,  away from the centre of the plate, such that $z = A\cos(2\pi r/\lambda)$.  Finally, a flat acrylic plate was used for control experiments. All plates were smooth to the touch.  For each plate 10 cm high acrylic walls were glued to each edge, to make a waterproof container.  

\begin{table}
\small
\centering
 \caption{Amplitudes $A$ and wavelengths $\lambda$ of the corrugated substrates (wave plates) used to template crack patterns. Linear wave plates are simply numbered, while radial wave plates are indicated by the suffix r.}
 \label{allvals}
\begin{tabular*}{0.48\textwidth}{@{\extracolsep{\fill}}cccc}
\hline
Plate & $A$ (cm) & $\lambda$ (cm) & $a$ = $A$/ $\lambda$ \\   \hline 
  1 & 0.25 & 2  &     0.125 \\
  2 & 0.25 & 1  &     0.25    \\   
  3 & 0.5 & 2   &   0.25        \\      
  4  & 0.25 & 0.5  &     0.5             \\
  5  & 0.5 & 1  &     0.5             \\ \hline
  1r & 0.25    & 1 & 0.25 \\
  2r & 0.25    & 0.5 & 0.5 \\  \hline
\end{tabular*}
\end{table}

Slurries of potassium bentonite (Acros Organics K-10 montmorillonite) were prepared by mixing a set weight of bentonite powder with twice as much water, by mass. Slurries were then stirred until smooth and immediately poured evenly into a container, over a substrate.  At this concentration the slurries behaved as fluids, rather than pastes.  These conditions avoid the memory effect \cite{Nakahara2005,Matsuo2012}, whereby the flow or vibration of a yield-stress paste can affect the way it cracks (see Fig. \ref{IntroFig}(f)).  We checked that this effect was absent by drying slurries over flat substrates, and finding isotropic crack patterns.

After pouring, the bentonite settled slightly, separating into a dense clay covered by a thin water layer.  We measured the thicknesses of these submerged clay layers in flat-bottomed containers after 6 hours of settling, without significant evaporation, and found that the consolidated wet clay consistently contained 0.49 grams of bentonite per cubic centimetre.  For slurries poured over the sinusoidal substrates we therefore used the total mass of the bentonite as a proxy for the average layer thickness $H$ of the wet clay layer.  All experiments were performed in the range of $H$ from 3 to 15 mm.

Once a slurry had been poured, it was left to dry in a quiet, warm environment for 6-48 hours, depending on the amount of clay used.  When drying is complete, the bentonite becomes lighter in colour; no further cracks appear after this time. For the linear wave plates, drying was accelerated by halogen heat lamps, which raised the surface temperature of the clay to $\sim50^\circ$C.  As elevated temperatures would melt the printed plates, drying over the radial plates was enhanced by raising the room temperature to about 30$^\circ$C.  During drying, the appearance and evolution of the cracks were imaged every 5 minutes using a digital SLR camera (Nikon D5000 series), mounted above the drying container.  By comparing the initial wet height $H$ to the dry height of various layers, we estimate a total vertical strain of 0.27, during drying.  

\subsection{Image processing} \label{improc}

\begin{figure}
\centering
\includegraphics[width =  8.3 cm]{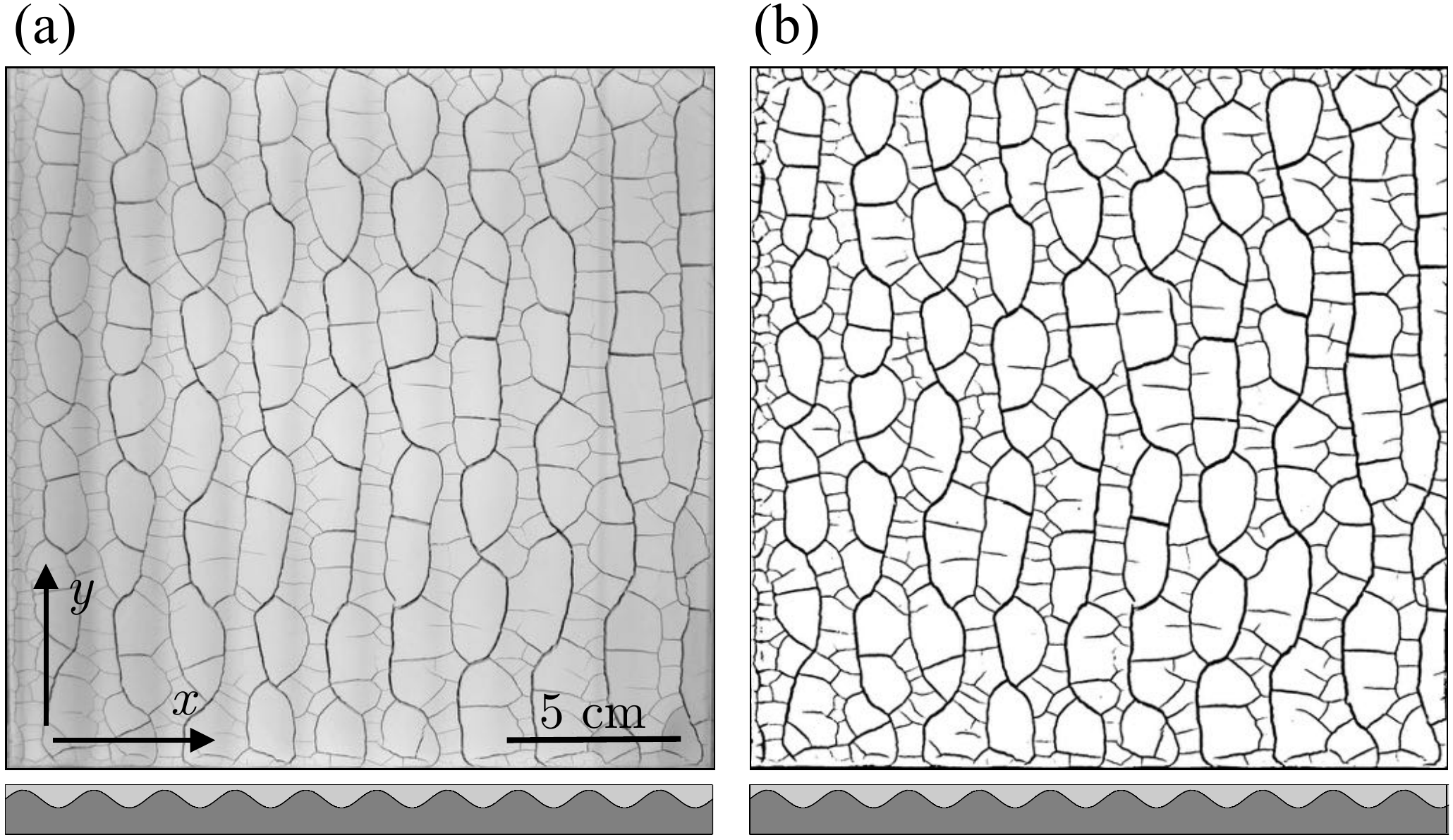}
 \caption{Image processing.  Each image (a) is cropped, contrast-adjusted, filtered and thresholded to (b) yield a black-and-white description of the crack positions.  The shape of the substrate (plate 1) is sketched below the images.} \label{rawnbin}
\end{figure}

Images were pre-processed using ImageJ and Matlab, as shown in Fig. \ref{rawnbin}. Each image was cropped to isolate the cracked clay,  then converted to grayscale by averaging across all colour channels, and contrast-enhanced.  Next, a band-pass filter was applied to remove high-frequency noise (3-pixel cutoff to reduce single-pixel noise) and low-frequency variations in lighting intensity (40-pixel cutoff, comparable to the width of the largest cracks).  Finally, each image was thresholded to give a black-and-white representation of the crack network, as in Fig.~\ref{rawnbin}(b).  

To characterise the crack patterns we will focus on the shape of the crack network, and on the relative positions and orientations of features such as the vertices between cracks.  For some measurements, we thus chose to eliminate unconnected cracks and spurs from the binary images.  Isolated cracks, and other noise, were removed by deleting objects below a threshold size.   Every image was then thickened, by 3 pixels, to ensure that small breaks in the crack network, or \textit{en-passant} cracks \cite{Fender2010}, were filled in.  The result was then eroded to leave a skeleton consisting of one-pixel-wide cracks that connected various vertices or nodes, and which delimited isolated patches, or \textit{peds}.  Cracks that branched off the main network, but stopped without connecting to anything, were then removed.    Both the binary images and the skeletons were used as the basis for further analyses.

Finally, we note here that the error bars given throughout this work are estimated by subdividing the processed images into quarters, analysing each subsection separately, and then calculating the standard deviation of the four different results.  

\section{Results and Discussion} 

\label{sectRes}
\begin{figure}
\centering
\includegraphics[width =  8.3 cm]{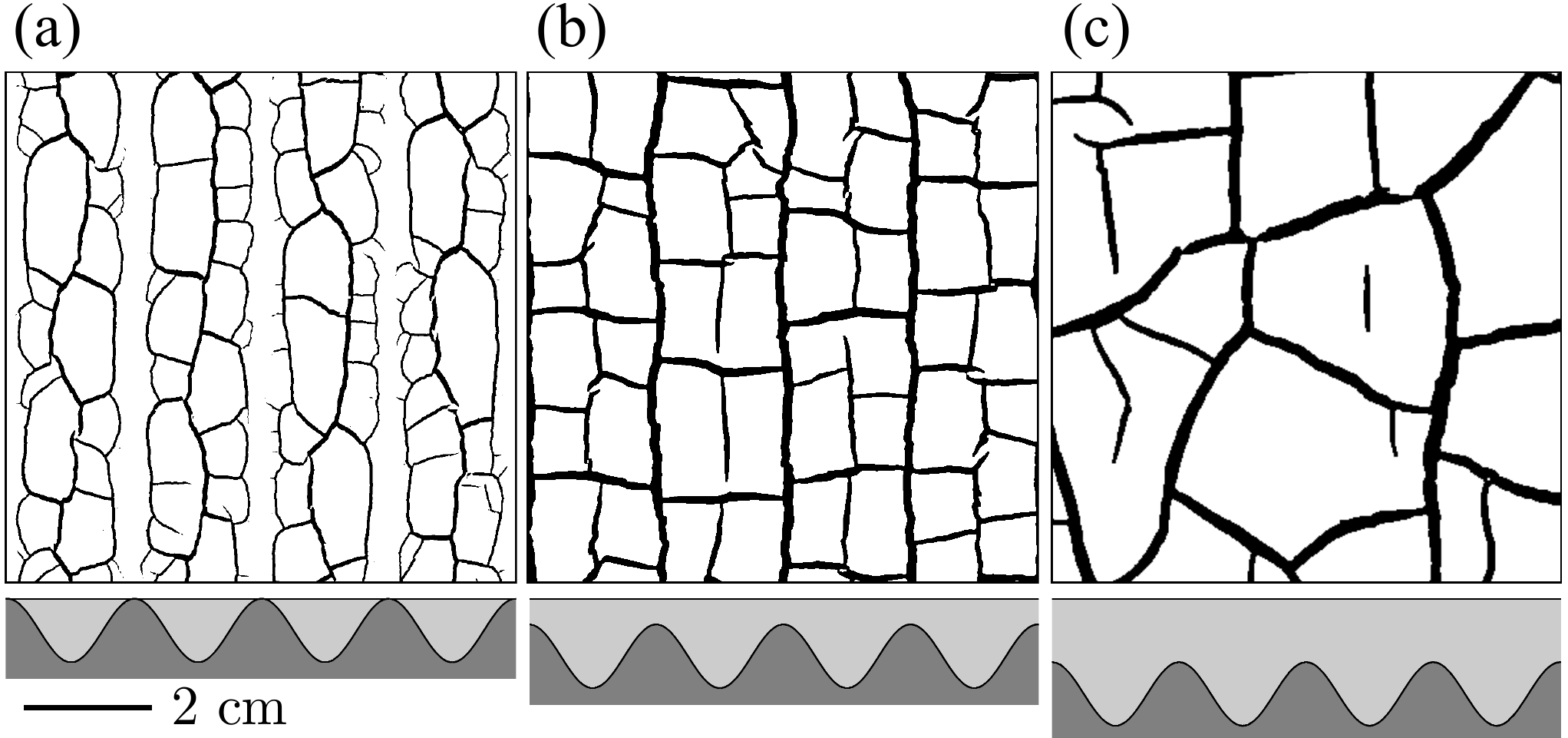}
\caption{Evolution of crack patterns: (a) for thin layers, wavy cracks are seen; (b) as the layer height increases straight cracks are observed, with laddering cracks between them; and (c) for thick layers the cracks are isotropic.  The substrate (plate 3) and clay thicknesses are sketched below each panel.} \label{ctrans}
\end{figure}

We found that as the thickness of the brittle layer increases, its contraction cracks slowly lose information about their substrate. However, in order to understand the effect of an uneven substrate, we must first briefly summarise the base case of cracks in flat bentonite layers.  For the flat control plate, isotropic crack patterns were seen for all layer thicknesses ($H$ from 4-15 mm).  Cracks appeared one by one, sequentially breaking the clay into smaller and smaller pieces, as in \cite{Bohn2005a,Bohn2005b}.  This process stops at some finite spacing, when the pattern saturates \cite{Bai2000,Goehring2015} -- further drying increases the widths of the cracks in these `mature' networks, as in \cite{Mal2005,Mal2007}, rather than create new ones.  We confirmed that the natural crack spacing over the flat plate is between 2-3 times the wet layer thickness $H$, using the line-dropping method \cite{Goehring2010}.  The results are compatible with the crack spacings reported in \cite{Goehring2010}, but consistently smaller, as the dry heights of bentonite layers were reported in that study.

For clays dried over the sinusoidally varying substrates the typical sequence of crack patterns, after drying is complete, is shown in Fig. \ref{ctrans}.  A wavy pattern developed in thin layers, with the first cracks meandering back and forth as they travelled along the troughs of the underlying substrate.  Various shorter cracks then filled in the pattern, as in Fig. \ref{ctrans}(a). For intermediate thicknesses of clay regular straight cracks appeared over each ridge, followed by cracks between these primary cracks, and at right angles to them, like the rungs of a ladder.  As in Fig. \ref{ctrans}(b), additional short cracks also often formed along the troughs of the pattern, especially in the later stages of drying.  For thick layers, as in Fig. \ref{ctrans}(c), there was no obvious effect of the substrate on the crack pattern, which was essentially isotropic.  

Linear elastic fracture mechanics has no inherent length-scales, other than those set by the geometry of the system under study.  If this assumption holds, then the dependence of the fracture pattern on the shape of our substrates can be entirely captured by considering two dimensionless parameters, such as the dimensionless layer thickness, $h = H/\lambda$, and a dimensionless amplitude, $a = A/\lambda$, of the substrate's relief.  Since in many situations, including ours, the spacing of thin-film cracks scales with the thickness of the cracking layer \cite{Allain1995,Shorlin2000,Bai2000,Kim2013}, the parameter $h$ should describe the relative match between the natural spacing of any cracks and the wavelength of the perturbations affecting their pattern.  The dimensionless amplitude $a$ describes the sharpness of the peaks on the substrate.   

\begin{figure}
\centering
\includegraphics[width =  8.3 cm]{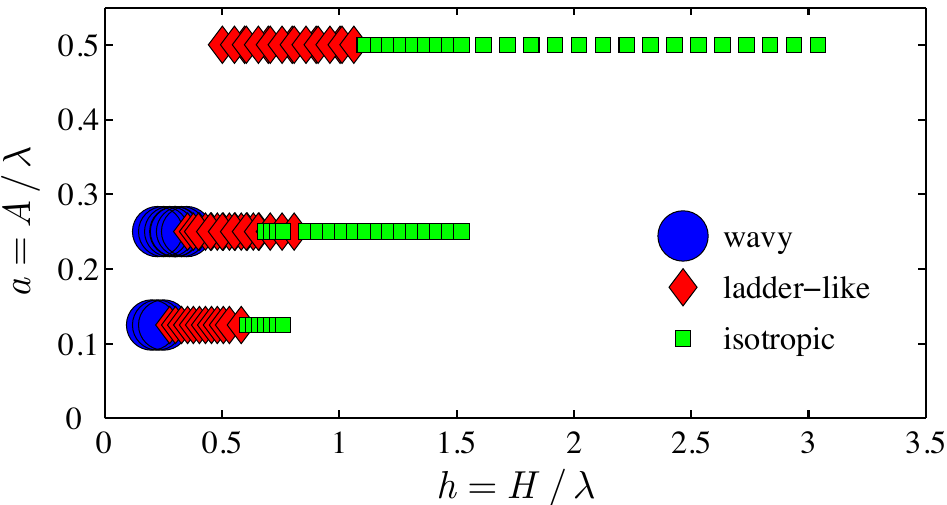}
\caption{\label{phasespace} A phase diagram showing where wavy, ladder-like and isotropic crack patterns were seen, for the linear wave plates.} 
\end{figure}

For the linear wave plates we categorised each fully-dried crack pattern by visual inspection as either wavy, straight or isotropic.  As summarised in Fig. \ref{phasespace}, the different patterns divide up the expected dimensionless phase space into well-defined regions.  At the boundaries between the different patterns were small areas of coexistence, for example between wavy and straight cracks.  This coexistence represents two scenarios -- either both patterns were observed within different regions of a single experiment, or were seen for the same $h$, but on different plates.  For dimensionless layer heights in excess of about $h=1$, only isotropic patterns were ever seen.  

Although there are clear visual differences between the patterns shown in Fig. \ref{ctrans}, there are no well-established methods by which they can be characterised.  In order to analyse our patterns quantitatively we therefore need to develop new order parameters similar to, for example, the orientational order parameter of liquid crystals \cite{deGennes1993}.  We present three such measurements next, based on Fourier methods, the relative orientation of the regions outlined by the crack network and the positions of neighbouring crack vertices.

\subsection{Fourier analysis of crack patterns}
  
\begin{figure}
\centering
\includegraphics[width =  8.3 cm]{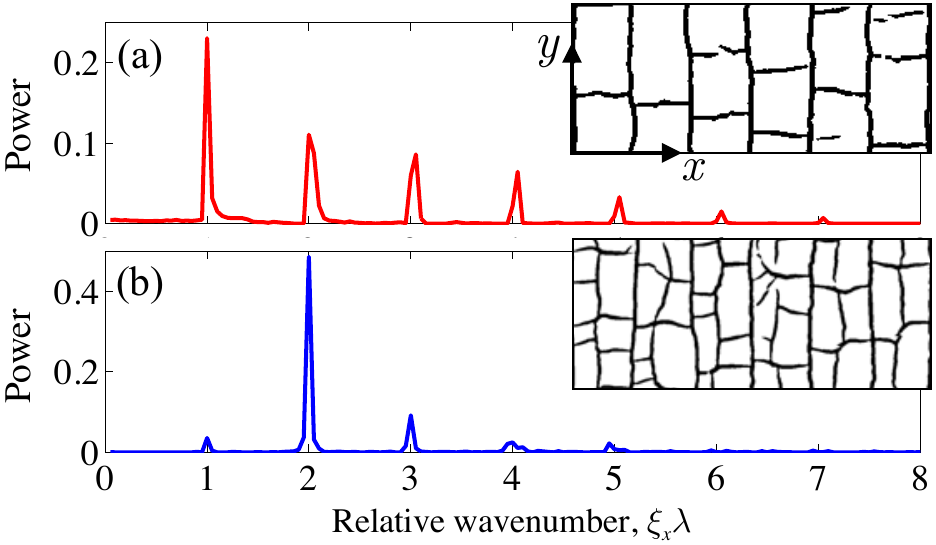}
 \caption{Example power spectra of ladder-like crack networks, with (a) cracks over each ridge of the substrate and (b) cracks over the ridges and troughs.  Insets show representative sub-sections, 6 cm wide, of the original images that were summed, vertically, to give average crack densities along the $x$-axis, for Fourier processing.}
\label{Fourier1}
\end{figure}

The presence of evenly spaced cracks that are well-aligned with the substrate's features suggests the use of Fourier methods to quantify our crack patterns.  Although a two-dimensional transform could be used to measure crack spacings, we focus here on a one-dimensional approach, which allows us to measure the relative alignment and periodicity of the cracks with respect to the linear substrates.  To this end, we first collapse each binary image to find the average density of cracks at each position along the $x$-direction.  This is done by summing over the $y$-direction of the images, where cracks are represented by ones and uncracked material by zeros.  We then find the power spectrum of the crack density as the absolute value of the square of its Fourier transform, normalised to have an integrated power of one.  

In cases where cracks are aligned with the substrate there are sharp peaks in the power spectrum  of the crack density when the wavenumber $\xi_x$ is an integer multiple of the wavelength $\lambda$ of the substrate, as in Fig.~\ref{Fourier1}.   In other words, the substrate prefers there to be a integer number of cracks, per wavelength.  The responses of all five linear wave plates were very similar: typically, most power was found in the first or the second peak, \textit{i.e.} when $\xi_x\lambda = 1$~or~$2$.  Furthermore, there was a change in the dominant mode of the power spectra around $h = 0.5-0.75$.  Below this, most power was in the $\xi_x\lambda = 2$ peak, while above this the $\xi_x\lambda = 1$ peak was also strong.  Visually, the point where these intensities cross over coincides with the transition between patterns with cracks along every ridge (Fig. \ref{Fourier1}(a), $\xi_x\lambda = 1$), and patterns with cracks along both the ridges and troughs (Fig. \ref{Fourier1}(b), $\xi_x\lambda = 2$).  The appearance of wavy cracks, at very small $h$, are associated with a decline in the power of the $\xi_x\lambda = 2$ mode.  Since they meander, their crack density is not as strongly localised as the ladder-like cracks, and the Fourier analysis reflects this. The isotropic cracks, around $h = 1$ or thicker, show no significant coupling to their substrate's periodicity.

\begin{figure}
 \centering
\includegraphics[width =  8.3 cm]{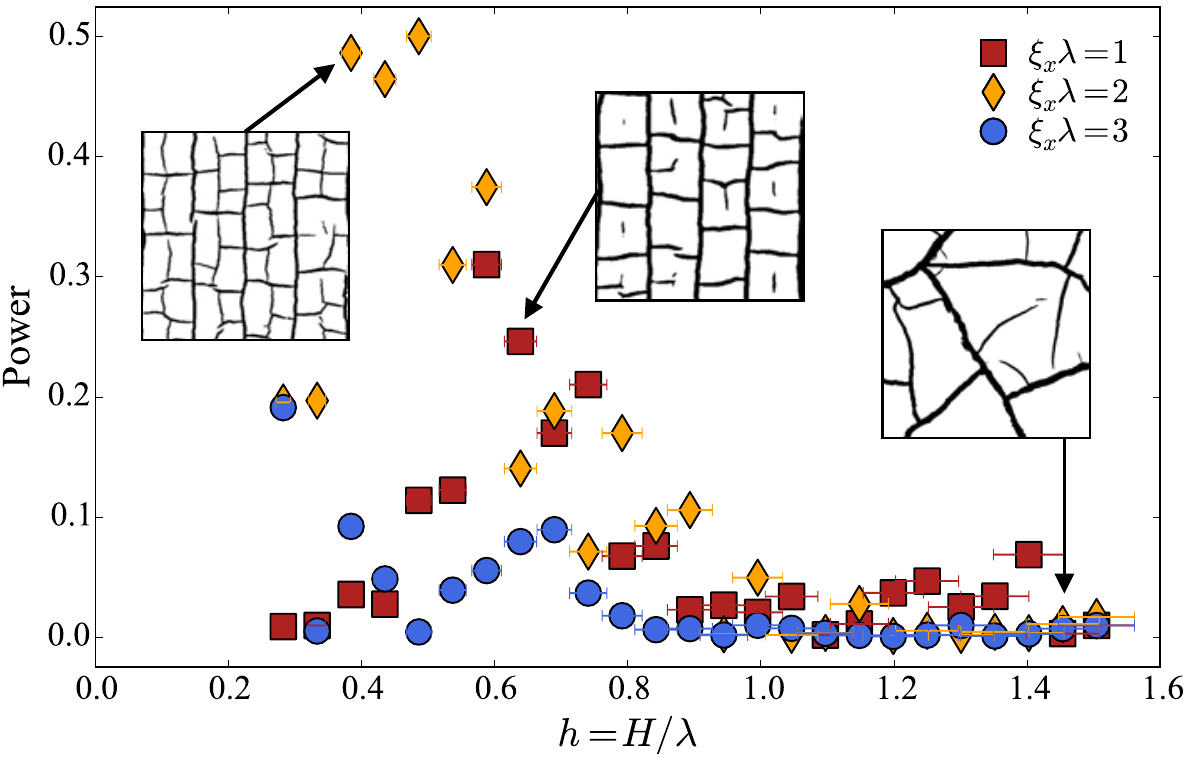}
 \caption{Spectral power of the lowest-order peaks for cracks over plate 2.  At $h = 0.6$ there is a transition from a pattern with cracks along the peaks and troughs of the substrate, to one with cracks only along the peaks.  Insets show 4$\times$4 cm$^2$ samples of crack patterns from the indicated conditions.}
\label{Fourier2}
\end{figure}

An example of all of the above behaviour is given in Fig.~\ref{Fourier2} for plate 2, where $a = 0.25$.  The results for plate 3 are consistent with this, and extend the range of suppressed $\xi_x\lambda = 2$ down to $h = 0.2$.  For plates 4 and 5, where the substrate is relatively sharper ($a = 0.5$), the sequence of responses is identical, but shifted to slightly higher values of $h$: the transition from one to two cracks per wavelength is at $h = 0.75$, for example.  For plate 1, with shallow relief, this transition is shifted down to about 0.4-0.5, but the Fourier analysis does not produce very clear results, otherwise.   These variations in the conditions leading to the different types of crack patterns are consistent with the visual classifications given in Fig.~\ref{phasespace}.

Thus, Fourier methods capture the regularity of the observed crack patterns, and quantify how the transition from one to two cracks per wavelength occurs when the periodicity of the substrate is about 1.5 to 2 times the thickness of the cracking layer -- \textit{i.e.} approaching the natural crack spacing of the clay over a flat substrate.  However, it does not provide a good indication of the alignment of the various crack features with the substrate.  The absolute intensity of the power spectral peaks, which could be such a measure, is too easily affected by noise to serve in this way.  The next two sections aim to provide more robust measures of alignment.  

\subsection{An orientational order parameter}

As the clay dries in our experiments, its cracks subdivide and isolate individual regions of material, whose shapes are then determined by their sets of final bounding cracks.   For example, in Fig.~\ref{ctrans}(b) the cracks around the dried peds are mostly parallel or perpendicular to the substrate relief. The cracked regions therefore have rectangular shapes, and tile the plane like rows of bricks. In contrast, the cracks in Fig.~\ref{ctrans}(c) are only weakly influenced by the substrate and define cracked regions with more arbitrary shapes and orientations.  This section aims to measure the degree of alignment of these regions, with respect to the ripples along their substrate.  

\begin{figure}
 \centering
\includegraphics[width =  8.3 cm]{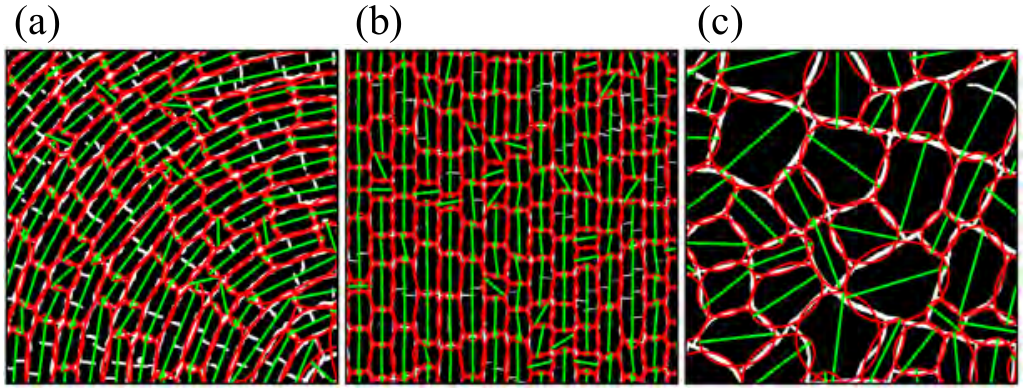}
 \caption{Ordering of cracked regions.  Ellipses are fit to each connected region outlined by cracks, and the orientational ordering of the ellipses is then studied.  Shown are the best-fit ellipses (red) with their major axes (green), overlaid on binary (white cracks on black background) representations of (a) well-ordered regions over a radial plate and (b) well-ordered and (c) isotropic regions over a linear plate.}
\label{nematic1}
\end{figure}

From each image we extracted a binary skeleton of the crack network (see Sect. \ref{improc}).  Each connected region in this image is identified and an ellipse is fit to cover its area.  The major axis of this ellipse goes through the region's centroid and its orientation, $\theta$, is chosen to minimise the second moment of area of the region, around the major axis.  Figure~\ref{nematic1} shows examples of elliptical fits of cracked regions with three different patterns.  

To characterise the crack-bounded regions, through their best-fit ellipses, we use an orientational order parameter that measures the relative alignment of rod-like particles, or any objects that are invariant after rotation by 180$^\circ$.  This parameter is well-known in the physics of liquid crystals (\textit{e.g} \cite{deGennes1993}), where it is used to describe the transition from an isotropic to a nematic (orientational order without positional order) phase, for example.  In our case, for dimension $d = 2$, the general definition \cite{Mercurieva1992} of this nematic order parameter simplifies to
\begin{equation}
S_1 = \frac{d\langle\cos^2\theta\rangle-1}{d-1} = \langle\cos(2\theta)\rangle,
\label{S1}
\end{equation}
where the angled brackets represent averaging over all ellipses.  The orientation, $\theta$, of the major axis of each ellipse is measured with respect to the $y$-axis for the linear wave plates, and the radial direction for the radially-symmetric substrates.  

 \begin{figure}
 \centering
\includegraphics[width =  8.3 cm]{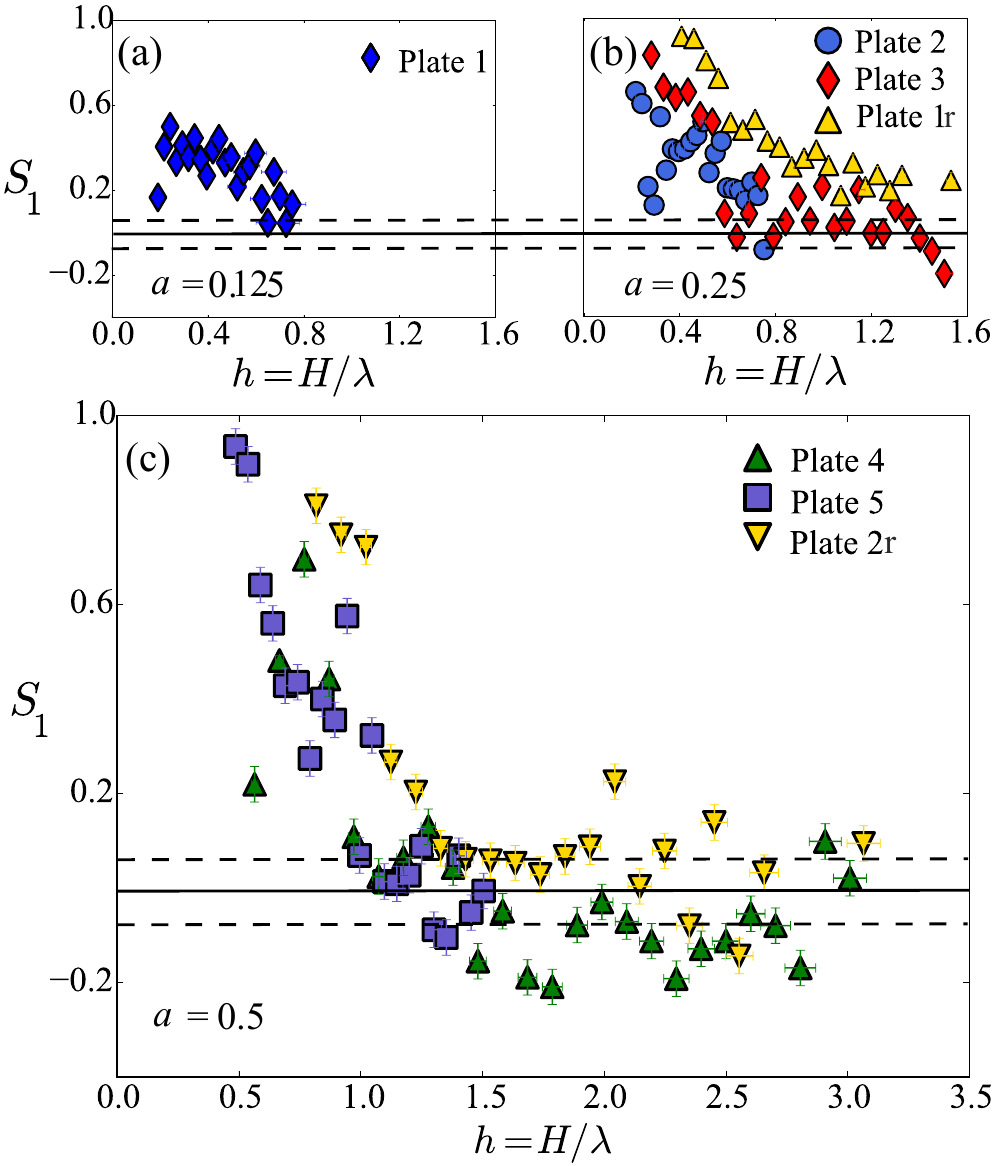}
 \caption{The orientational order parameter $S_1$ (Eq. \ref{S1}) describes how well-aligned individual peds (continuous regions, entirely bounded by cracks) are with the orientation of their substrate.  Results are given for wave plates with dimensionless amplitudes (a) $a = 0.125$, (b) $a = 0.25$ and (c) $a = 0.5$.  The solid and dashed lines represent the mean and standard deviation of measurements of $S_1$ on flat plates.}
\label{nematic2}
\end{figure}

The orientational ordering for all our experiments is shown in Fig. \ref{nematic2}.  For each dimensionless amplitude $a$ there is good agreement between the different experiments, although the radial plates give slightly higher values of $S_1$ than the linear plates.  For small enough $h$, in particular around $h = 0.5$ and below, the peds are very strongly aligned with the relief of the substrate, and this alignment slowly decreases as the layer thickness increases, until $S_1$ becomes statistically equivalent to its value over the flat plate above a relative thickness of at most $h = 1.25$.  Furthermore, the intensity of the ordering decreases with the relative sharpness of the plates, $a$, for experiments with the same $h$.  However, unlike the Fourier analysis, or the order parameter we introduce in the next section to describe the relative alignment of the individual cracks, $S_1$ remains high for very low $h$.  It is a sensitive measure of the straight-to-isotropic transition, but because it averages over the entire shape of a cracked region, it does not detect the meandering of the cracks.

\subsection{Crack alignment order parameter}

A second order parameter is introduced to describe the relative alignment of the cracks with their substrate.   Working from the skeletons of the cracks, we first find the coordinates of all vertices, or intersections between cracks.  These skeletons form closed networks, where all isolated features and spurs have been removed.  The vertices are therefore points along the skeleton that touch at least three different cracked regions, and this property is used to identify them.  Crack \textit{segments} are then defined by the paths between any two neighbouring, connected vertices, and their orientation $\phi$ by the vectors connecting these two endpoints.   Since one end of a crack cannot be effectively distinguished from the other, the distribution of crack angles is collapsed onto the range from 0--180$^\circ$.  These angles are referenced to the $y$ or radial directions, for the linear or radial wave plates, respectively.    

\begin{figure}
\centering
\includegraphics[width =  8.3 cm]{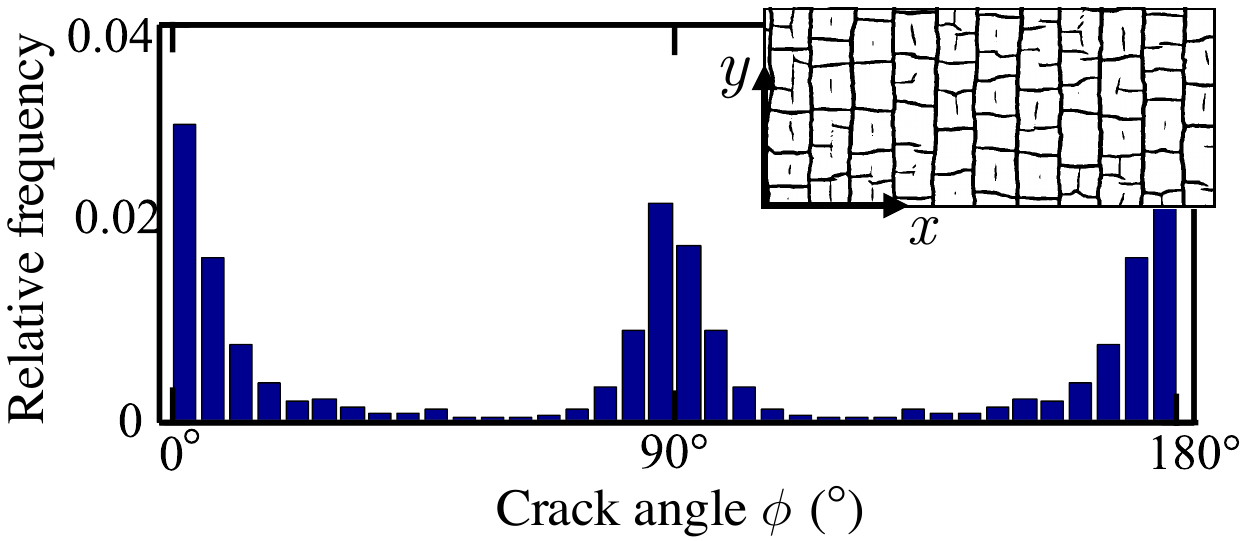}
 \caption{The distribution of crack orientations in a ladder-like pattern.  The peak at 0$^\circ$ and 180$^\circ$ (these two angles being equivalent) reflects the dominant cracks parallel to the $y$-axis, while the  peak at 90$^\circ$ comes from the shorter `rungs' of the ladders.}
\label{angles1}
\end{figure}

It should be emphasised that this method creates a very local definition of the orientation of a crack, and splits long cracks into many individual segments, between adjacent intersections with other cracks.  This is different from some other network analyses of cracks \cite{Andresen2013,Bohn2005b,Hafver2014}, which focus instead on the entire length of a crack, regardless of any intersections it has with other features. An example of a crack pattern, and the relative abundance of crack orientations in it, is given in Fig.~\ref{angles1}.

\begin{figure}
\centering
\includegraphics[width =  8.3 cm]{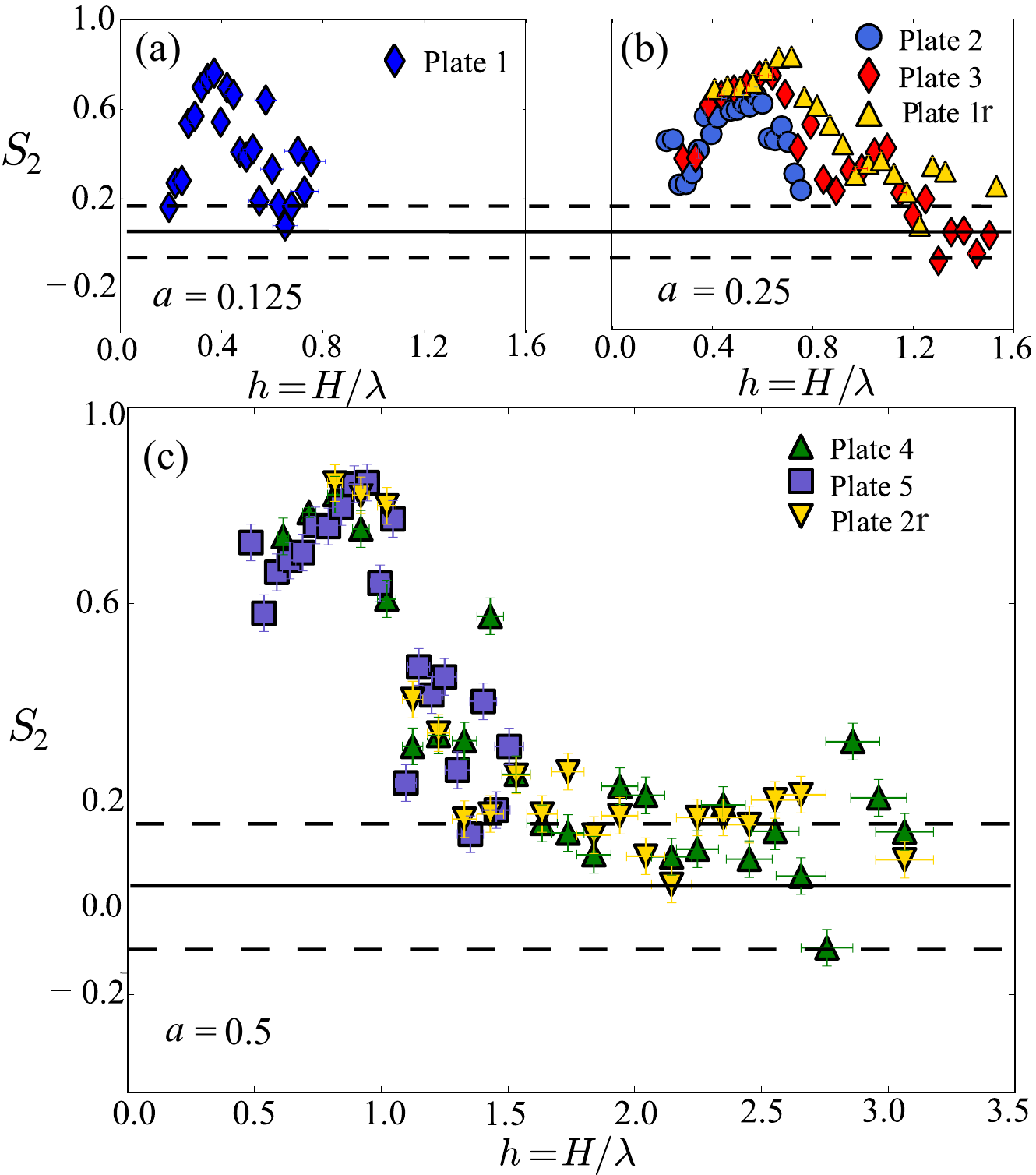}
 \caption{The crack alignment order parameter $S_2$ (Eq. \ref{S2}) measures the relative alignment of the cracks with the substrate, and takes large values for networks with cracks that are parallel or perpendicular to the underlying relief.  Shown are the results for (a) $a = 0.125$, (b) $a = 0.25$, and (c) $a = 0.5$.  The solid and dashed lines represent the mean and standard deviation of $S_2$ for the flat plate.}
\label{angles2}
\end{figure}

In our more well-ordered experiments, as in Fig. \ref{angles1}, there are typically prominent cracks along the ridges of the substrate, and shorter perpendicular cracks that join these together.  Their measure must reflect this symmetry.  In analogy with $S_1$, it should take a value of $+1$ for cracks that are either parallel or perpendicular to some reference direction, $0$ for random patterns, and $-1$ for cracks that are all completely \textit{mis}-aligned with the reference (here, cracks at $\pm$45$^\circ$ to the substrate features).  The simplest such definition is for a crack alignment order parameter
  \begin{equation}
S_2 = \langle \cos(4\phi) \rangle.
\label{S2}
\end{equation}
We calculated this parameter for each drying experiment, and summarise the results in Fig.~\ref{angles2}.

We found that the crack alignment order parameter, $S_2$, matches the orientational order parameter of the cracked regions, $S_1$, in most situations.  There is also good data collapse between the linear and radial wave plates, and between all plates of the same relative amplitude $a$, confirming that our results do not depend on the absolute scale of the drying experiments.  There is a loss of ordering around $h = 1$, again at slightly higher values of $h$ for the sharper substrate ridges, and slightly lower values for the shallower relief.  However, unlike $S_1$, there is a clear downturn in $S_2$ at low layer thicknesses -- the cracks become less aligned for very thin layers.   As it is a local measure, the crack orientation is more sensitive than our other metrics to any curvature of the cracks, and some of the decrease in $S_2$ for very low $h$ is due to the wavy cracks.  Nevertheless, in all cases $S_2$ starts to fall well before we can see wavy cracks by eye.   In Section \ref{Smodel} we will show how this effect, and the other transitions in the crack pattern, can be explained by the relative forcing of the uneven substrate, which will be shown to be strongest at the intermediate layer thicknesses where both order parameters are largest.  

\subsection{Extension to other crack networks}

The two order parameters $S_1$ and $S_2$ introduced above describe, respectively, the relative alignment of crack-bounded areas, or individual cracks, with respect to some known angle. In our experiments there was an obvious direction to choose for this angle, based on the substrates used -- either the local radial vector, or one of the cartesian axes.  However these metrics can also be used to determine whether a crack network is aligned or not without knowing, in advance, any such preferred direction.  By extending the liquid-crystal analogy of the order parameter $S_1$, one may use the local \textit{director}, or average local orientation of the relevant objects, as the reference direction.  In other words, one may first calculate the average $\theta$ or $\phi$, either over all cracks or areas in a region of interest, or locally over neighbourhoods that are large enough to provide good statistical averaging.   Then, in calculating the order parameters through Eqs. \ref{S1} and \ref{S2}, all angles can be referenced to this local director orientation.  This would allow one to compare the relative degree of alignment of natural crack patterns like polygonal terrain \cite{Sletten2003,Pina2008} and the hydraulic channels of heavily fractured rock \cite{Valentini2007,Andresen2013}, for example.  It would also allow one to better measure manufactured patterns such as craquelure\cite{Bucklow1998,Crisologo2011}, or custom-designed micro-cracks \cite{Nam2012,Kim2013}.

\section{Modelling the crack pattern transitions}
\label{Smodel}

When a crack opens, it releases stored strain energy from its environment.  In Griffith's model of fracture this energy is used to create the new surface area of the crack, as it grows \cite{Griffith1921,Lawn1993}.  In this section we consider a finite element model of a channel crack opening over a sinusoidal surface, and use the Griffith criteria for fracture to predict where the first cracks will appear.  We will show how the observed transitions from wavy to straight to isotropic cracks can be explained in terms of the relative strain energy released by a crack opening at different locations in a thin elastic layer drying  -- or cooling, or with some other misfit strain --  over such a rippled substrate. 

In particular, we consider a linear poroelastic \cite{Biot1941} model for cracks in drying clay.  The total stress $\bar\sigma$ in the clay is a sum of the elastic stress $\sigma$ supported by the network of clay particles and the capillary pressure $p$ of the interstitial liquid, such that 
\begin{equation}
\bar\sigma_{ij} = \sigma_{ij} - p\delta_{ij}.
\label{poro1}
\end{equation}  
The negative sign here accounts for the convention that a positive stress is tensile, while a positive pressure is compressive.  The pressure is generated by the tiny curved menisci of the water still trapped in the clay, and for these conditions $p<0$.  The network stress can be given in terms of the strain $\epsilon$ by the linear elastic constitutive relationship 
\begin{equation}
\sigma_{ij} = \frac{E}{1+\nu}\bigg(\epsilon_{ij} + \frac{\nu}{1-2\nu}\epsilon_{ll}\delta_{ij}\bigg),
\label{poro2}
\end{equation} 
where $E$ is the Young's modulus and $\nu$ the Poisson's ratio of the clay. The strain tensor is, in turn, derived from the elastic displacements, $u$, of the clay by
\begin{equation}
\epsilon_{ij} = (\partial_iu_j + \partial_ju_i)/2.
\label{poro3}
\end{equation}
The pressure term in Eq. \ref{poro1} provides the driving force for fracture.  As the clay dries $p$ becomes more and more negative (its magnitude increases), requiring the clay to compensate with a tensile strain that can lead to crack growth. In order to instead model cooling, or tissue growth (\textit{e.g.} for crocodile skin \cite{Milinkovitch2013} or leaves \cite{Laguna2008}), this pressure can be replaced by an equivalent thermal contraction term \cite{Biot1956,Norris1992}, or the effects of differential growth in a brittle layer and its substrate.  

\begin{figure}
\centering
\includegraphics[width =  8.3 cm]{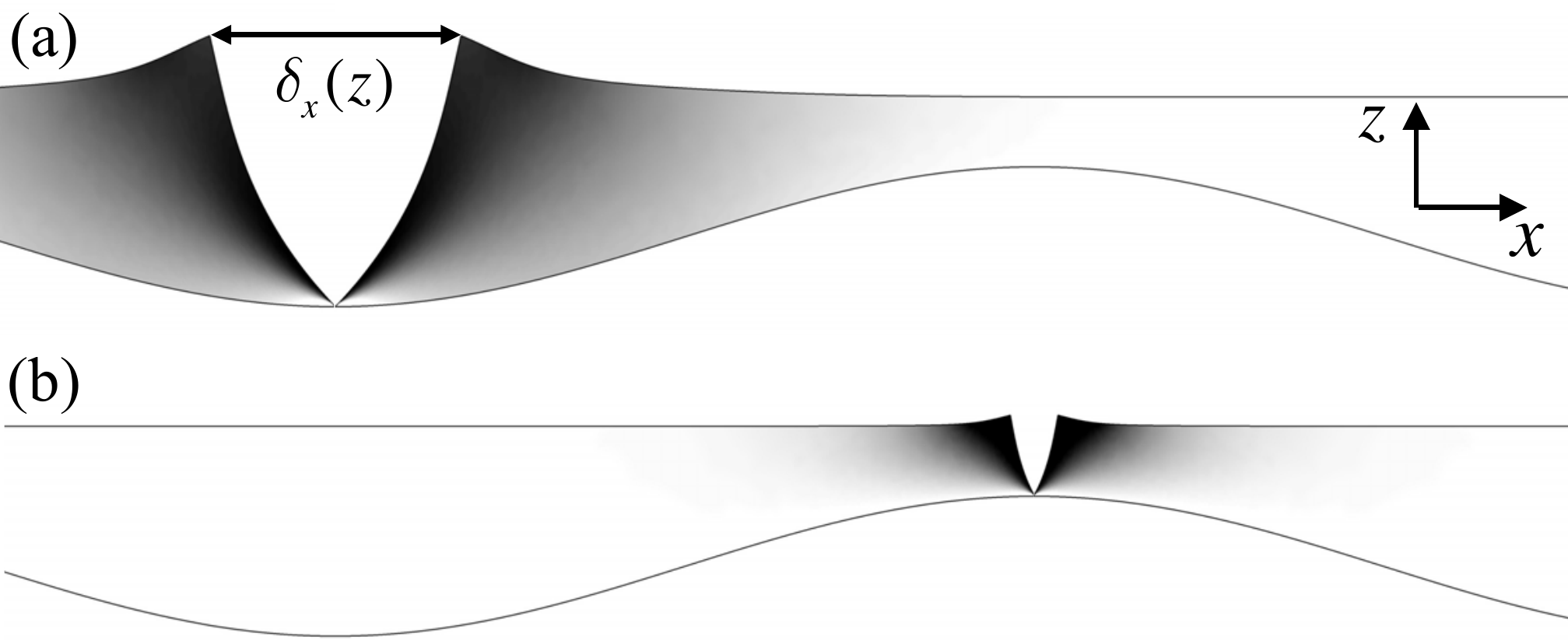}
 \caption{Numerical solution of a channel crack opening over (a) a trough or (b) a peak of the substrate.  The deformations caused by the cracks are exaggerated, so as to be visible.  The greyscale shows how much of the pre-crack stress $\sigma_{xx}^*$ is released by the crack -- with black shown where all the stress is released, and white where the crack does not affect the stress at all.}
\label{model1}
\end{figure}

We solved Eqs. \ref{poro1}--\ref{poro3} using Matlab's finite-element routines, on domains like those shown in Fig. \ref{model1}, namely semi-rectangular regions with straight edges at the top ($z = 1$, dimensionless units, scaled so that the average film thickness $H=1$), left ($x = 0$) and right ($x = 10$) sides, and a corrugated base where $z = 1 - \ell = A\cos(2\pi x/\lambda + \psi)$.  Here $\psi$ is an offset phase, allowing us to examine the strain energy released by fracture at various positions along the substrate, and $\ell$ is the local film thickness.   Plane strain conditions, with no displacements allowed in the $y$-direction, were used to model a channel crack advancing with a constant profile along the sinusoidal plate.   The upper surface was taken to be traction free ($\sigma_{xz} = \sigma_{zz} = 0$ on $z=1$), while the lower surface was treated as no-slip, such that $u = 0$ there, and only vertical displacements were allowed ($u_x = 0$) along the right boundary.  The left-face boundary was either  treated like the right one, for an intact film, or as a traction-free boundary, for a cracked film.   Finally, the pressure $p$ was taken to be constant everywhere, as appropriate for quasi-static fracture.  

The work done by the growth of the channeling crack can be measured by integrating the product of the pre-crack stress, $\sigma^*$, and the opening displacements of the crack, $\delta$, as in \cite{Lawn1993,Beuth1992}.  This is, essentially, the sum of the forces on the incipient crack surfaces, acting over the distance they travel as the crack opens. The strain energy released, per unit area of new crack, is thus
\begin{equation}
G = \frac{1}{2\ell} \int_{1-\ell}^1 \delta_x \sigma_{xx}^* + \delta_z \sigma_{xz}^* dz.
\end{equation}
For a simulation with a specific $A$ and $\lambda$ we calculated the pre-crack stress field once, using $\psi = 0$, and used this solution to find $\sigma^*$ at different positions along the substrate.  We then calculated the opening displacements of a crack at different $\psi$.  Accounting for the opening displacement to either side of a crack can be done by letting $\delta_x = u_x(\psi) + u_x(-\psi)$, and $\delta_y = u_y(\psi) - u_y(-\psi)$.  This gives us the values of $G$ for cracks growing along the substrate at different $\psi$, ranging from the crest of the sine wave, to its trough.  

For a flat substrate, this problem has been studied in depth by Beuth \cite{Beuth1992}, Xia and Hutchinson \cite{Xia2000} and Yin \textit{et al.} \cite{Yin2008,Yin2010}.  They showed that $G$ scales as
\begin{equation}
\label{Gscale}
G = \frac{\pi}{2}\frac{\sigma_0^2\ell g}{\bar{E}},
\end{equation}
where $g$ is a dimensionless parameter that only depends on the elastic mismatch between the cracking film and its substrate, and where $\bar{E} = E/(1-\nu^2)$ and $\sigma_0 = -p(1-2\nu)/(1-\nu)$.  For example\cite{Beuth1992}, for $\nu = 1/3$, $g = 0.71$. We confirmed their solutions with respect to the exponential decay of the height-averaged stress away from the crack \cite{Xia2000} and the dependence of $g$ on $\nu$ \cite{Beuth1992} and the depth of fracture \cite{Beuth1992,Yin2008}, and checked that $g$ was independent of the average film thickness, the film's elastic modulus $E$ and the capillary pressure $p$.  The mesh used, and the horizontal domain length, were found to be sufficient to agree with these known solutions to within 1\%.

\begin{figure}
\centering
\includegraphics[width =  8.3 cm]{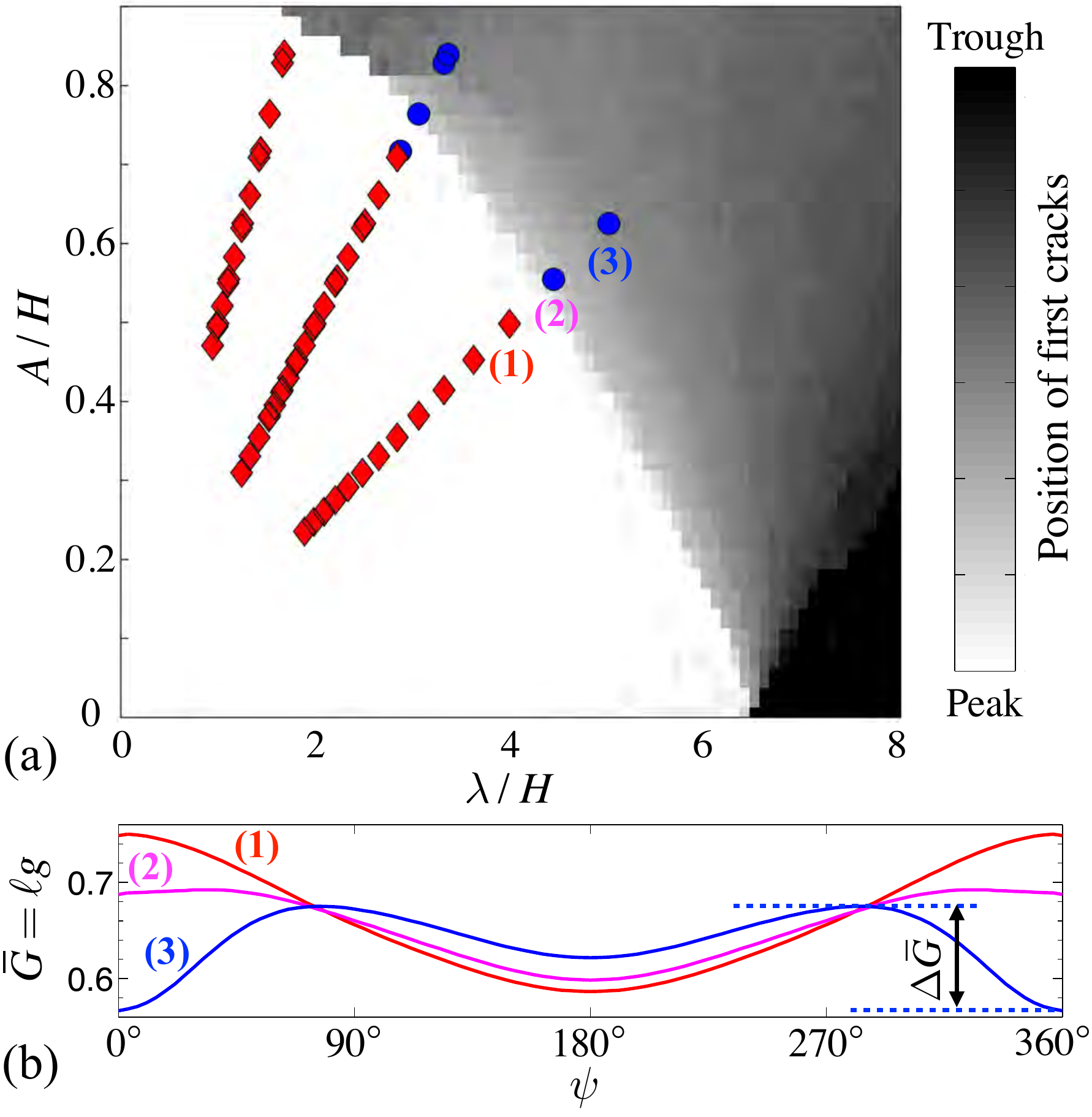}
\caption{ Modelling the straight-to-wavy transition.  (a) Predicted positions of the first cracks, based on the offset $\psi$ where the energy release rate $G$ is maximised. Red diamonds and blue circles show the experimental parameters where ladder-like or wavy cracks were seen, respectively (from Fig. \ref{phasespace}). (b)~Energy landscapes of $\bar{G}(\phi$) for three conditions, indicated by (1-3) in (a), which cross over the straight-to-wavy transition.  Under these conditions there is a bifurcation from one energy maximum for a crack on the top of a ridge, to two preferred locations, on either side of the ridge.}
\label{model2}
\end{figure}

For an uneven substrate $G$ depends on location, measured here by the offset $\psi$.  As the clay dries the relative shape of $G(\psi)$ will not change, but rather its magnitude will simply increase along with the capillary pressure.  The Griffith criteria predict that fractures can only grow once $G$ reaches some critical value, $G_c$, which is a material parameter (see \textit{e.g.} \cite{Lawn1993}).  The first cracks in the brittle drying layer should therefore form at the positions where the strain energy released by these cracks, \textit{i.e} $G$, would be a maximum, as these are the first places where $G = G_c$.   For a Poisson ratio of $\nu = 1/3$, we calculated how $G$ varied with position in a wide range of experimental conditions, from $A = 0.01$--0.90 and $\lambda = 0.5$--8.0.  To focus only on geometric effects we used Eq. \ref{Gscale} to rescale $G$ by the relative stiffness $2\bar{E}/\pi\sigma_0^2$, so that $\bar{G} = \ell g$. The results are summarised in Fig. \ref{model2}(a), which shows the values of $\psi$ that maximise $G$ for various $A$ and $\lambda$.  Depending on the shape of the substrate and the thickness of the clay, the preferred positions of the earliest cracks can be either (i) over the peaks of the substrate, (ii) over the troughs or (iii) at pairs of points displaced to either side of each peak.  

\begin{figure}
\centering
\includegraphics[width =  8.3 cm]{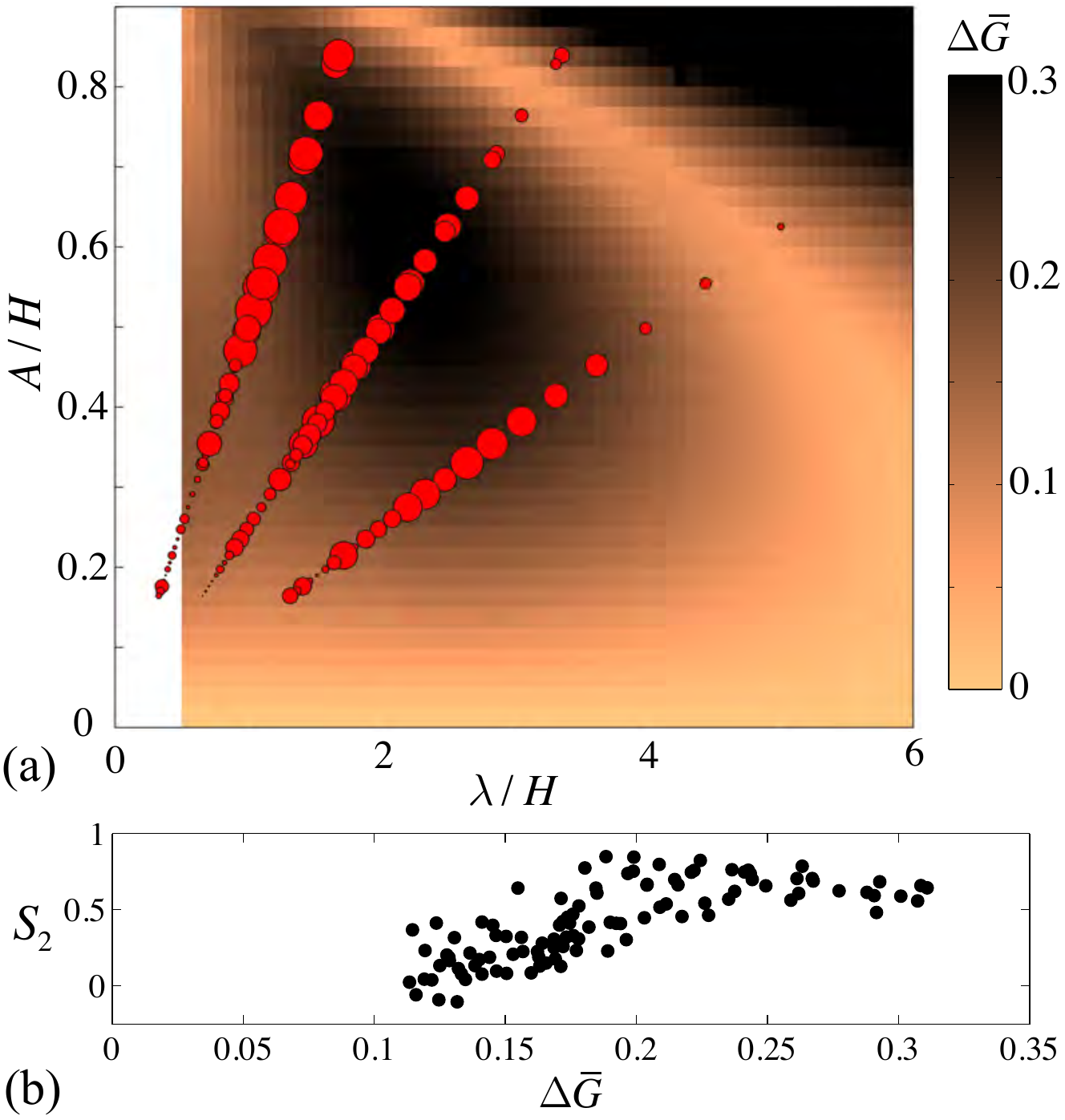}
\caption{Modelling the straight-to-isotropic transition.  For (a) the colour indicates the difference between the maximum and minimum values of $\bar{G}$.  The red circles have a radius proportional to the crack orientation order parameter, $S_2$, for each experiment.  The most well-ordered patterns correspond to the patch of high $\Delta\bar{G}$ around the centre of the image. Panel (b) shows how $S_2$ is well-correlated with $\Delta\bar{G}$, between experiments where straight or isotropic patterns were seen, and corresponding simulations.}
\label{model3}
\end{figure}

Figure \ref{model2}(a) also replots the observations from Fig. \ref{phasespace}, corresponding to the conditions where wavy and ladder-like cracks are found.  For all experiments where straight cracks were seen the model correctly predicts that the first cracks should appear over the ridges of the substrate.  Where wavy cracks are seen, the model instead generally predicts two stable positions for cracks, along the sides of the ridges.  Although our model has no dynamic component, this empirical agreement of the division of phase-space suggests that the wavy cracks result from cracks moving back and forth between their two bistable positions.   The boundary between the straight and wavy cases corresponds to a forward pitchfork bifurcation, as sketched in Fig. \ref{model2}(b).    

To consider the more gradual transition from ladder-like to isotropic cracks, we looked at the relative variations in $\bar{G}$, defining $\Delta\bar{G}$ as the difference between the maximum and minimum values of $\bar{G}(\psi)$, for each value of $A$ and $\lambda$ studied.   If $\Delta\bar{G}$ is small, then there is only weak guidance for the cracks, and they should appear more random than in cases of larger variation of energy release rate.  We tested this hypothesis by comparing $\Delta\bar{G}$ with the crack alignment order parameter of our experiments.  First, we calculated $\Delta\bar{G}$ for each point in our simulation, mapping the results in Fig. \ref{model3} alongside the data showing $S_2$ for our various experiments with linear wave plates.  This figure shows that there is a neighbourhood of driving parameters, centred around $\lambda/H = 2.2$ and $A/H = 0.6$, where $\Delta\bar{G}$ is largest.  Our experiments pass through this region, and show their most well-ordered patterns as they do so.  To quantify this agreement, we then calculated $\Delta\bar{G}$ for the conditions corresponding to each experiment where ladder-like or isotropic cracks were seen.  As shown in Fig. \ref{model3}(b), there is indeed a good correlation between the range of energy release rates allowed for any particular geometry, and the relative alignment of the cracks that formed over it.  

Finally, we note that some insight into the limits of our results can be gained through a qualitative comparison to channel cracks in flat films and coatings. Here the stresses relaxed by the crack will decay away over a length comparable to the thickness of the film \cite{Hutchinson1992,Xia2000}.  This effect can also be seen in Fig. \ref{model1}. For a wavy substrate this limit is approached when the amplitude of the relief goes to zero, and under such conditions isotropic cracks with a spacing $\sim H$ are clearly expected.  However, for finite $A/H$ the limit can still offer insight.    For very long wavelength ripples, large $\lambda/H$, each crack will see a local environment, of size $\sim H$, that is approximately flat.  Although these conditions were not probed by our experiments, we would also expect an isotropic pattern in this case, but one with a local modulation of the crack spacing reflecting the changes in local thickness.  For the opposite limit, of very sharp features where $\lambda/H\ll 1$, the region around any crack will sample many ridges and troughs, and these local variations will tend to cancel each other out (see Saint-Venant's principle \textit{e.g.} \cite{Sadd2005}); again, isotropic cracks would be expected.  In other words, for finite-amplitude relief a brittle layer must filter out the high and low-frequency information of its substrate, and any contraction cracks that form should reflect this.

\section{Conclusions}

Our results explore the effects of an uneven substrate on the development of contraction cracks in thin films, and are inspired by natural patterns of cracks around buried craters, paintings, and vapour-deposited thin films.  We showed that the wavy, periodic undulation of a substrate can cause the ordering of a crack pattern that forms above it, and has the strongest influence on this pattern when the thickness of the cracking layer is comparable to, or slightly smaller than, the periodicity of the substrate.    

Generally we found that for increasing layer thickness the crack pattern changed from one with wavy cracks running along the troughs of the substrate, to straight cracks appearing over each peak and trough of the substrate, to straight cracks on only the peaks of the substrate, and finally to a disordered or isotropic pattern.  The wavy-to-straight transition was the most abrupt change, as evidenced by examples of single experiments that were wavy on one side of the plate, and straight on the other.   All these patterns were quantified by Fourier methods and orientational order parameters, similar to the nematic order parameter of liquid crystals, which are easily adaptable to other types of crack networks.   Finally, we showed how the transitions between pattern type could be explained by a simple finite-element model of channeling cracks advancing over uneven relief, regardless of whether those cracks were driven by drying, cooling or the differential growth of crocodilian skin.  

\acknowledgements

We thank U. Schminke and M. Wolff for assistance with the design and manufacture of our plates; LG thanks P. Keen and A. Chetwood for their contributions to an early version of this experiment; PN thanks A. Fourri\`ere for helpful discussions.

\providecommand*{\mcitethebibliography}{\thebibliography}
\csname @ifundefined\endcsname{endmcitethebibliography}
{\let\endmcitethebibliography\endthebibliography}{}

\end{document}